\documentclass[12pt]{amsart}
\usepackage{geometry} % see geometry.pdf on how to lay out the page. There's lots.
\usepackage{graphicx}
\usepackage{amssymb}
\begin{document}

{A Natural Seismic Isolating System: The Buried Mangrove Effects}
\\
\\

Philippe Gueguen$^1$, Mickael Langlais$^1$, Pierre Foray$^2$, Christophe Rousseau$^2$ and Julie Maury$^1$\\

1 ISTerre/CNRS/LCPC, University Joseph Fourier, Grenoble\\

2 L3SR, Institut National Polytechnique INP, Grenoble\\
\\
\\
\\

Accepted (12/2010) for publication in Bulletin of seismological Society of America.\\
\\
\\

Revised version: 21th of December 2010\\

\newpage

ABSTRACT\\

The Belleplaine test site, located in the island of Guadeloupe (French Lesser Antilles) includes a 3 accelerometers vertical array, designed for liquefaction studies. The seismic response of the soil column at the test site is computed using three methods: the spectral ratio method using the vertical array data, a numerical method using the geotechnical properties of the soil column and an operative modal analysis method (Frequency Domain Decomposition).The Belleplaine test site is characterized by a mangrove layer overlaid by a stiff sandy deposit. This configuration is widely found at the border-coast of the Caribbean region, which is exposed to high seismic hazard.  We show that the buried mangrove layer plays the role of an isolation system equivalent to those usually employed in earthquake engineering aimed at reducing the seismic shear forces by reducing the internal stress within the structure. In our case, the flexibility of the mangrove layer reduces the distortion and the stress in the sandy upper layer, and consequently reduces the potential of liquefaction of the site.

\newpage

\section{Introduction}

The near surface geological condition is one of the critical factors in controlling the seismic ground motion variability and its amplification, with a high impact on the variability of the damage pattern from large earthquakes. Main contributions to the amplification of motion are the body wave trapping effects due to the impedance contrast between horizontally layered sediments and underlying bedrock for a 1D medium (Aki and Richards, 2002), and lateral trapping of surface waves in 2D and/or 3D geometries (Cornou et al., 2003). In both cases, site effects lead to a frequency-dependent amplification of the seismic ground motion. Site effects analysis are primarily carried out on surface recordings (Borchedt, 1970; Lachet et al.,1996; Gueguen et al., 1998; 2000). Another option for estimating the soil response is to use vertical arrays which has provided advances in the understanding of shallow layer seismic responses, including non-linear behavior of the soil column (Satoh et al., 1995; 2001), downgoing waves producing destructive interferences (Bonilla et al., 2002) and wave attenuation in sediments (Archuleta et al., 1992;  Abercrombie, 1997). Soil profiles are mostly characterized by shear-wave velocities increasing with depth but irregular velocity profile, which occur more infrequently, may provide peculiar site responses with consequences on the seismic ground motion estimates and on the post-seismic damage pattern. This is the case of sea shore regions in Caribbean islands often characterized by reverse velocity profiles such as the Guadeloupe Island (French lesser Antilles) (e.g., Gagnepain et al., 1995; Roulle and Bernardi, 2010). These site effects are mainly due to the presence of mangrove swamps filled with limestone from the surrounding hills (Roulle and Bernardi, 2010). \\
We have shown that the presence of a buried mangrove layer plays the role of a seismic isolation device by reducing the seismic deformation in the upper and stiffer layer. Passive seismic isolation techniques are usually employed to reduce the deformation in the building by adding a soft layer between the soil and the building, most often made of rubber bearings. In the case of a stiff structure with respect to the isolating system, the building oscillates as a rigid body at the natural frequency of the bearings (Buckle and Mayes, 1990). With suitable isolating systems, the seismic deformation produced by the horizontal seismic ground motion is supported by the rubber bearings, implying deformations rather limited in the structure. \\
To improve the assessment of the seismic deformation of the soil column, we applied a non parametric Operative Modal Analysis (OMA), which consists of extracting physical parameters of the system using in-situ recordings without any assumption on the model. Such techniques, commonly employed by the engineering community, are used in order to characterize the dynamic response of a system and to detect changes due to non-linearity by comparing the shape of the physical modes (He and Fu, 2001; Carden and Fanning, 2004; Cunha and Caetano, 2005). The main goal of this paper is to compare the seismic response of the soil column at the Belleplaine test site obtained by a standard method based on spectral ratios with the OMA method, and to show the usefulness of the modal analysis method for detecting the seismic deformation and non-linear effects during earthquakes. \\
After describing the vertical array and the geotechnical cross section of the Belleplaine test site, the seismic response using the spectral ratio technique is analyzed. Modal analysis is then performed using earthquake data recorded in the borehole and the experimental mode shape is finally compared to linear and nonlinear 1D modal responses of the soil profile.      
 
\section{The Belleplaine experimental site}
The Belleplaine vertical array test-site is located in the Guadeloupe Island (French Antilles) close to the Caribbean subduction zone (Figure 1A). The site was designed in the framework of the Belleplaine French National project (ANR-06-CATT-003) for liquefaction analysis in the case of sea shore sediment materials, including extensive in-situ geotechnical and geophysical surveys (drilling boreholes and laboratory testing on sample, SASW, H/V seismic noise ratio survey, seismic piezocone), pore pressure measurements and accelerometric ground motion sensors. The velocity model at the top 35 meters is known from synthesis of borehole drillings and downhole seismic piezocone penetrometer (Santruckova, 2008; Foray et al., 2011) summarized in Figure 1B. Five cone penetration tests with additional pore pressure measurements (CPTU) were carried out using seismic piezocones penetrometers. As the aim of these tests was to quantify the properties of the superficial liquefiable layer, only the first 14m were investigated. The five penetration tests were located close to the two instrumented boreholes (50 m between the two accelerometric boreholes) and their results were remarkably similar, showing a good homogeneity of the stratigraphy of the site (Santruckova, 2008; Foray et al., 2011).
\\

The soil structure is composed of a shallow 1 m-thick layer with an S-wave velocity $\beta_1=200 m/s$, overlying a 4 m-thick stiff sandy layer ($\beta_2$=470 m/s) below which is found a soft and consolidated mangrove layer (33 m thick) with an S-wave velocity $\beta_3=220 m/s$. The bedrock is  GL-38m and it is characterized by reef coral limestone for which no S-wave velocity information are available. The vertical array is composed of three synchronized triaxial accelerometers (EST Shallow-Borehole Episensor) placed at GL-0m, GL-15m and GL-39m, where GL means "ground level" (Figure 1B). The GL-15m sensor is located within the mangrove layer, 10 m below the mangrove/sand interface. The GL-39m sensor is inserted in the bedrock layer, immediately under the mangrove/bedrock interface. \\
The set of records used in this study corresponds to local and regional events (Figure 1A), localized by the Guadeloupe Observatory (OVSG). It consists of recordings from 62 earthquakes, with $M_L$ between 2 and 6.4 and epicentral distance ranging between 20 and 450 km. During the installation, the horizontal components of the instruments placed at GL-15m and GL-39m deviate $85$ and $81$ degrees in clockwise direction, respectively, as estimated using long period seismic waves from the most distant event ($M_L$=6.4 at 450 km). Before analysis, the horizontal components are corrected, applying a rotation of $81^o$ and $85^o$ in the counterclockwise direction. Because of the high dynamic range of the acquisition system and the broad-band nature of the accelerometric sensors, no pre-processing algorithms are applied to the data, except for the offset correction. Figure 2 displays the accelerometric ground motion recorded by the vertical array at the three sensors for the most distant event: in the time domain, we clearly observe the amplification of the seismic ground motion due to the soil column, i.e. between the amplitudes at different depths. The maximal horizontal peak ground acceleration (PGA) recorded at GL-0m is 5 $cm/s^2$ which corresponds to a weak ground motion (Idriss, 1990), i.e. only linear seismic response is expected here. The comparison of the motion recorded at the surface and in depth (Figure 3) shows that horizontal PGA at the bottom (GL-39m) are approximately two times smaller than at the surface (GL-0m) and at intermediate depth (GL-15m), even though GL-15m sensor is at equal distance from the two others. We observe similar characteristics on the accelerometric response spectra Sa at 1s period (Figure 3, upper row). The response spectra confirm that the accelerometric ground motion is mainly controlled by the ground motion at 1s, while at lower frequency (i.e., Sa at 3s period, Figure 3, middle row) the ground motion is not modified by the soil profile.  
 
\section{Processing}
We analyze the seismic response of the soil column using the spectral ratio method (Aguirre and Irikura, 1997). For all the data, we select three portions of the accelerograms: (1) a 60-sec window containing the P and S wave part of the record, (2) a 10-sec window centered on the S-wave arrival time, and (3) a 50-sec window of the coda at the end of the record. Fast Fourier Spectra are computed for each horizontal component at the GL-0m, GL-15m and GL-39m stations. As suggested by Field and Jacobs [1995], the spectral ratios are computed for frequencies having a signal-to-noise ratio above 3. After smoothing the spectral amplitudes according to the Konno-Ohmachi window (Konno and Ohmachi, 1998) (b=30), the spectral ratios GL-0m/GL-39m, GL-0m/GL-15m and GL-15m/GL-39m are computed and averaged over all the recorded data, separately for the E-W and N-S components, and plotted with the standard deviation (Figure 4). Amplification is observed at 1.3 Hz (Figure 4) for the GL-0m/GL-39m and GL-15m/GL-39m spectral ratio, corresponding to the resonance frequency of the site. We also estimated the resonance frequency by the analytical Rayleigh method (Dobry et al., 1976). The fundamental frequency of the soil profile only depends on the S-wave velocity $\beta_3$ and thickness $H$ of the soft mangrove layer, derived from the oversimplified 1D relationship $f_o=\beta_3/4H (=1.5Hz)$.The 1D response assumption is also confirmed by comparable spectral ratios at the fundamental frequency of the soil column considering the two horizontal ground accelerations (Figure 4).  \\

Below 2 Hz, the GL-0m/GL-15m ratio (Figure 4) is approximately constant, also inferred by the comparable Sa values at 1s computed at  GL-0m and GL-15m (Figure 3). The ground motion at the fundamental frequency of the site (1.3Hz) is almost the same at GL-0m and GL-15m, the two levels moving in phase and with the same amplitude. This observation indicates that the soil column between GL-0m and GL-15m moves as a stiff layer, without any internal deformation.  At 2.3 Hz, the spectral ratio is influenced by the downgoing wave effects. This phenomenon was studied by several authors (e.g., Bonilla et al., 2002; Aguirre and  Irikura, 1997): at any depth, the ground motion contains the incident waves and the wave reflected at the surface, with opposite phase for same frequencies. The result is a destructive interaction of body waves producing a hole in the spectral ratio curve. \\

To better understand the observation, we compute the 1D theoretical transfer function using the R/T coefficient method (Kennett, 1974). As the quality factors in the sediments ($Qp$ and $Qs$) are not directly known, we used the oversimplified empirical relationships $Qs=\beta (m/s)/10$ and $Qp=2Qs$ (e.g.  Olsen et al., 2003;  Abercrombie, 1997). While the 1D resonance frequency is only a function of the thickness and of the S-wave velocity in sediments, the amplification factor is dependent on the S-wave velocity contrast, i. e. in our case the contrast between the soft mangrove layer and the deep limestone bedrock (GL-38). As the S-wave velocity in the bedrock $\beta_4$ is not directly known, we tested three S-wave velocity for this medium: 1000m/s, 1500 m/s and 2000 m/s. Once the transfer function computed at GL-0m and GL-15m, we convolved all the GL-39m horizontal recordings with the calculated transfer functions computed at GL-0m (GL-0m) and GL-15m (GL-15m) receivers and the same spectral ratio procedure is applied as for the observed data. Because of the likeness of the North-South to the East-West component at the fundamental frequency of the soil profile, only spectral ratios observed in the North-South direction are displayed in Figure 5. The synthetic spectral ratios reproduce well not only the 1.3 Hz frequency peak values but also its amplitude using the $\beta_4=$1500 m/s assumption. For this value of $\beta_4$, the synthetic ratio reproduces both the GL-0m/GL-39m and GL-15m/GL-39m frequency peak around 1.3 Hz and the amplification factor. In addition, the downgoing effect is also reproduced on the GL-0m/GL-15m ratios, with some frequency shift of the inverted peak, and some differences between data and synthetic spectral ratios above 2Hz. The misfit at high frequencies and at 2.3 Hz may be due to the smoothed velocity profile inferred from seismic piezocone penetrometer. Indeed, the smoothing of the soil profile does not account for the transition zone of the properties between each layer.      

\section{Modal analysis of the soil column}

For stratified 1D soil profiles, simplified procedures exist for estimating the fundamental period and mode shape of a linear model of soil profiles (Dobry et al., 1976). One approximate method is based on the modal analysis considering the soil profile as a continuous shear-beam. As only the first mode of the system is discussed here, the Rayleigh procedure is used in this paper. This algorithm is based on the exact solution obtained by equalizing the total maximum kinetic and potential energies of the free-oscillating response of the system at the fundamental mode. By introducing the equilibrium between inertia and elastic forces at any level $z$, the equation of the first mode is given by the equation (1): 

\begin{equation}
 X(z)=\int^{z}_{0} \frac{dz_i} {\rho(z_i)\beta^2(z_i)} \int^{H}_{z_i} \rho(z_{i+1})X(z_{i+1})dz_{i+1}
\label{Eq1}
 \end{equation}

where $X(z)$=$X(z_{i+1})$ is the first mode shape, $z$ the depth of the interface between layer $i$ and layer $i+1$, $H$ the thickness of layer $i$, $\beta$ the S-wave velocity and $\rho$ the density, and $i$ and $i+1$ the layer indexes under the assumption that the first mode shape of the soil column is composed of $n$ cosine curves, one for each layer. The origin of the coordinate axis $z$ is defined at the bottom of the deposit, i.e. at the GL-38m level in our case. An approximate solution of the Rayleigh solution can also be used, considering a constant density with depth ($\rho(z)=\rho$). The resulting expression derived from Equation (1) becomes:

 \begin{equation}
 X_{i+1}=X_i + \frac{H-z_{mi}}{\beta_i^2}H_i
\label{Eq2}
 \end{equation}

where $X_i$ and $X_{i+1}$ are the estimate of the fundamental mode shape at lower and upper boundaries of layer $i$, $H_i$ the thickness of layer $i$ and $H-z_{im}$ the depth of the middle of layer $i$. 

Recently, several new algorithms have been provided from engineering community for processing ambient vibrations with operational modal analysis finalities (He and Fu, 2001; Carden and Fanning, 2004; Cunha and Caetano, 2005). Among them, the Frequency Domain Decomposition (FDD) method was selected in this study (Brincker et al., 2001; Michel et al., 2008; Michel et al., 2010). As it is a non-parametric method, no a priori model is needed for processing the data. It allows the estimate of the eigenvalues of the system (mode shapes and frequencies) by diagonalizing the Power Spectra Density (PSD) matrix, i.e. by computing the Fourier spectra of the cross-correlation matrix obtained by simultaneous recordings done in the system. Brincker et al. (2001) showed that the PSD can be decomposed into singular vectors and scalar singular values. Ventura et al. (2003) and Michel et al. (2008) also applied this method with success using earthquake data recorded in building under the assumption of white noise spectra in the frequency range of the seismic data.  

% the estimate of the output Pseudo Spectral Density $G_{yy}$ matrix of the responses $y(t)$ known at each frequency $\omega_i$ is decomposed by computing the Singular Value Decomposition (SVD) of the spectral matrix by $G_{yy}(j\omega_i)=U_iS_iU_i^H$, where $U_i$ is the unitary matrix composed by the singular vectors $u_{ij}$, and $S_i$ is the diagonal matrix of the scalar singular values $s_{ij}$. 

When a single mode is dominating, the first singular vector is an estimate of the mode shape $\phi$. At a resonance frequency, the first singular value exhibits a peak and the corresponding singular vector is an estimate of the mode shape. By comparing the mode shape at the peak to the mode shapes at neighboring frequencies, it is possible to select the bell of the mode in the singular values, using the Modal Assurance Criterion (MAC) (Allemang and Brown, 1982). \\

One possible way to estimate variations of the mechanical characteristics of the system (e.g. due to non-linear effects) is to compare the variations of the modal parameters. For example, Wu et al. (2009) showed a shift of the resonance frequency of a soil column during strong motion and Allemang and Brown (1982) discussed the variation of the shapes of the modes during non-linear processes. To identify if such variations may be present at the Belleplaine site, the experimental mode shape is computed for three sets of data, selected as a function of the value of the horizontal PGA (N-S and E-W components) at the GL-0m sensor (Figure 6). The small variations of the shape of the first mode versus the PGA range whatever the direction is in favor of an 1D elastic response assumption of the soil column, as mentioned in the previous sections. Moreover, we observe a very good fit between the two numerical estimates of the mode shape (exact and approximated Rayleigh method) and the median value of the mode using the three (i.e., very weak, weak and moderate ground motion) sets of data (Figure 6). As suggested also by the spectral ratio technique, the seismic response of the soil column is mainly controlled by the buried soft layer (mangrove), the deformation in the upper sandy layer being rather limited at this mode (i.e., at 1.3 Hz). \\

Ground motion at the Belleplaine site are clearly too weak to go to non-linear behavior. Consequently, the seismic modal analysis is only representative of the elastic domain. As recently and experimentally confirmed by Wu et al. (2009), the site response parameters (i.e., frequency and damping) may temporarily change as function of the level of the ground motion, and can be used to quantify the non-linear seismic response. Moreover, whatever the vibrating system, the shape of the modes is also sensitive to nonlinear processes and therefore sensitive to the damaging process within the system (e.g., Allemang and Brown, 1982). \\
We simulate the nonlinear seismic response of the Belleplaine test site using the Cyclic1D free software (Elgamal et al., 2002). The FEM model was developed to simulate the cyclic mobility response mechanism and the pattern of the shear strain accumulation in two-phases materials (solid-liquid). In our case, we considered a three layers soil profile, composed of a topmost stiff sand layer (8m thick), an intermediate-depth layer made of a cohesive and soft layer (clay layer, 32 m thick) overlying rigid bedrock. The input motion corresponds to a 0.2 g sinusoidal motion, having 10 cycles and applied at the bottom of the soil column. The accelerometric time histories are computed at a regular depth sampling (Figure 7A) and the modal analysis is performed using the synthetics and applying the FDD method described above (Figure 7B). The nonlinear effect implies a change in the shape of the fundamental mode, essentially due to the temporal change of the physical properties of the soil. As for the linear behavior, we observe that the majority of the distortion is supported by the intermediate-depth soft layer while the topmost layer has no clear internal distortion. This interpretation must be confirmed by future seismic strong ground motion recorded at the Belleplaine test site. However, we observe that experimental modal analysis may be of great interest for analyzing the effect of nonlinearity at depth and to assess the modal seismic response of a natural system.

\section{Discussion and conclusion}
Seismic base-isolation of structures has been applied throughout the world since the middle of the XXth century. One worldwide device (passive) consists of decoupling the structure from earthquake induced ground motion, via a flexible support intercalated between the ground and the structure. One classical system is composed by rubber elastomeric bearings which provide a reduction in the stiffness or spring constant between the structure and the ground. With suitable flexible supports, the main philosophy of such as a device is to provide only limited internal stress in the structure under severe shaking by shifting the frequency response away from the frequency of the maximal shaking energy. Such systems have two important functions (Buckle and Mayes, 1990): (1) the frequency of the isolated structure is decreased to a value beyond dominant frequencies for typical earthquakes and (2) the displacement is controlled by the addition of an appropriate amount of damping. An equivalent rheological model is usually proposed as a non-deformable mass resting on the ground through the flexible support. \\

The Belleplaine test site presents a comparable behavior. Since a very soft layer (mangrove) is overlain by stiff soil (sand), we are in the same configuration as the seismic base isolation system used for earthquake engineering. The seismic response of the soil column is mainly controlled by the properties of the soft layer (flexible support) which supports the maximum part of the seismic distortion. By this way, the strain and the distortion in the uppermost sandy layer (i.e. non-deformable mass) is rather limited. The ability of the sand layer to produce liquefaction is then reduced, even if the primarily in-situ investigation highlighted risk of liquefaction at the Belleplaine site. \\

Moreover, the advantage of the Belleplaine configuration is to reduce the variability of the maximal seismic ground motion since the maximal seismic energy is controlled by the seismic response of the soft mangrove layer. This observation may have many implications for reducing the effects of ground shaking on infrastructure. Firstly, a lot of sub-tropical regions exposed to seismic hazard have coasts constituted by a soil column similar to the Belleplaine site and are therefore expected to reduce the seismic risk as compared to standard estimates. Secondly, the seismic hazard at such sites may be controlled by the frequency response of the mangrove which is characterized by an almost constant value of the dominant frequency, independently of the frequency of the input motion. By reducing the variability (in frequency) of the seismic ground motion, one of the crucial point for predicting hazard and damage to structures, it should therefore be feasible to adapt a cheap building design by avoiding the resonance of the soil column. This point may be of great interest for countries located in the sub-tropical regions.  \\

The OMA approach used in this paper is relevant for (1) defining the deformation of the soil column during earthquakes and (2) detecting the variations of the mode shapes during strong motion due to non-linear processes. This technique, commonly employed by the engineering community, is used to characterize the dynamic response of a system and to detect changes due to non-linearity by comparing the shape of the physical modes. Experimental mode shapes may therefore give deeper insight to where, in depth, the maximum deformation would take place as well as potential locations of non-linear behavior, these two may not coincide. The experimental analysis requires "expensive" data provided by vertical array. The monitoring of the site could be designed for specific infrastructures sensitive to seismic non-linear effects. \\

\section{acknowledgments}

This work has been supported by the French Research National Agency (ANR) through Cattel program (project BELLEPLAINE n¡ANR-06-CATT-003). The installation of the accelerometric sensors and the access to the data were funded by the French Accelerometric Network (RAP). We thanks Annie Souriau, Nikos Theodulidis and Helle Pedersen for their fruitful comments.

\section{Data and Resources}
Accelerometric data used in this study were collected as part of the National Data Center of the French Accelerometric Network (RAP-NDC). Data can be obtained from the RAP-NDC at www-rap.obs.ujf-grenoble.fr.\\
Well logs were provided by the Belleplaine project of the French Research National Agency (ANR) through Cattel program.

\newpage

\section{Figure caption}

Figure 1: A) Epicenters of the earthquakes recorded at the Belleplaine test site and used in this study. B) Description of the soil profile and position of the accelerometric sensors.  \\

Figure 2: Example of N-S recordings at the three GL sensors  ($M_L$=6.4 at 80 km). \\

Figure 3: Seismic ground motion recorded at GL-39m (open squares) and at GL-15m (filled squares) as a function of surface ground motion (GL-0m) for  vertical (left), North-South (middle) and East-West (right) components. Upper row: Accelerometric response spectra at 1s - Middle row:  Accelerometric response spectra at 3s - Lower row: Peak ground acceleration (PGA). \\

Figure 4: Average spectral ratio of GL-0m/GL-39m (upper row), GL-0m/GL-15m (middle row) and GL-15/GL-39m (lower row) for N-S (right) and E-W (left) components for the frequencies with a Signal-to-Noise ratio over 3. The black continuous line shows the average ratio (+/-  standard deviation in dotted line) for the time window including P and S waves. Superimposed gray lines show the average ratio for time windows including only S waves or Coda waves.\\

Figure 5: Comparison between the observed North-South spectral ratio (black) and computed (red) at the vertical array (Belleplaine test site) using the Kennett (1974) method:  GL-0m/GL-39m (top, left), GL-0m/GL-15m (top, right)/GL-15/GL-39m (bottom, left), using three values of S-wave velocities in the bedrock layer.\\

Figure 6: Experimental first mode in the North-South (left) and East-West (right) directions, using the FDD method and for three ranges of peak ground acceleration (moderate: PGA$>0.35$ cm/s$^2$; weak: $0.15 <PGA<0.35$ cm/s$^2$; very weak: PGA$<0.15$ cm/s$^2$).\\

Figure 7: A) Synthetics of seismic ground motion in the Belleplaine soil profile using the Cyclic1D FEM software developed by (Elgamal et al., 2002). The red signal is the input motion corresponding to a 0.2 g sinusoidal motion, with 10 cycles and applied at the bottom of the soil column. B) Comparison between the average experimental first mode (N-S direction) with the  complete (1) and simplified (2) Rayleigh method, and with the nonlinear mode shape at the Belleplaine test site computed using Cyclic1D. \\

\newpage

%%%%%%%%%%%%%%%%%%%%%%%%%%%%%%%%%%%%
\begin{figure}
\noindent\includegraphics[width=40pc]{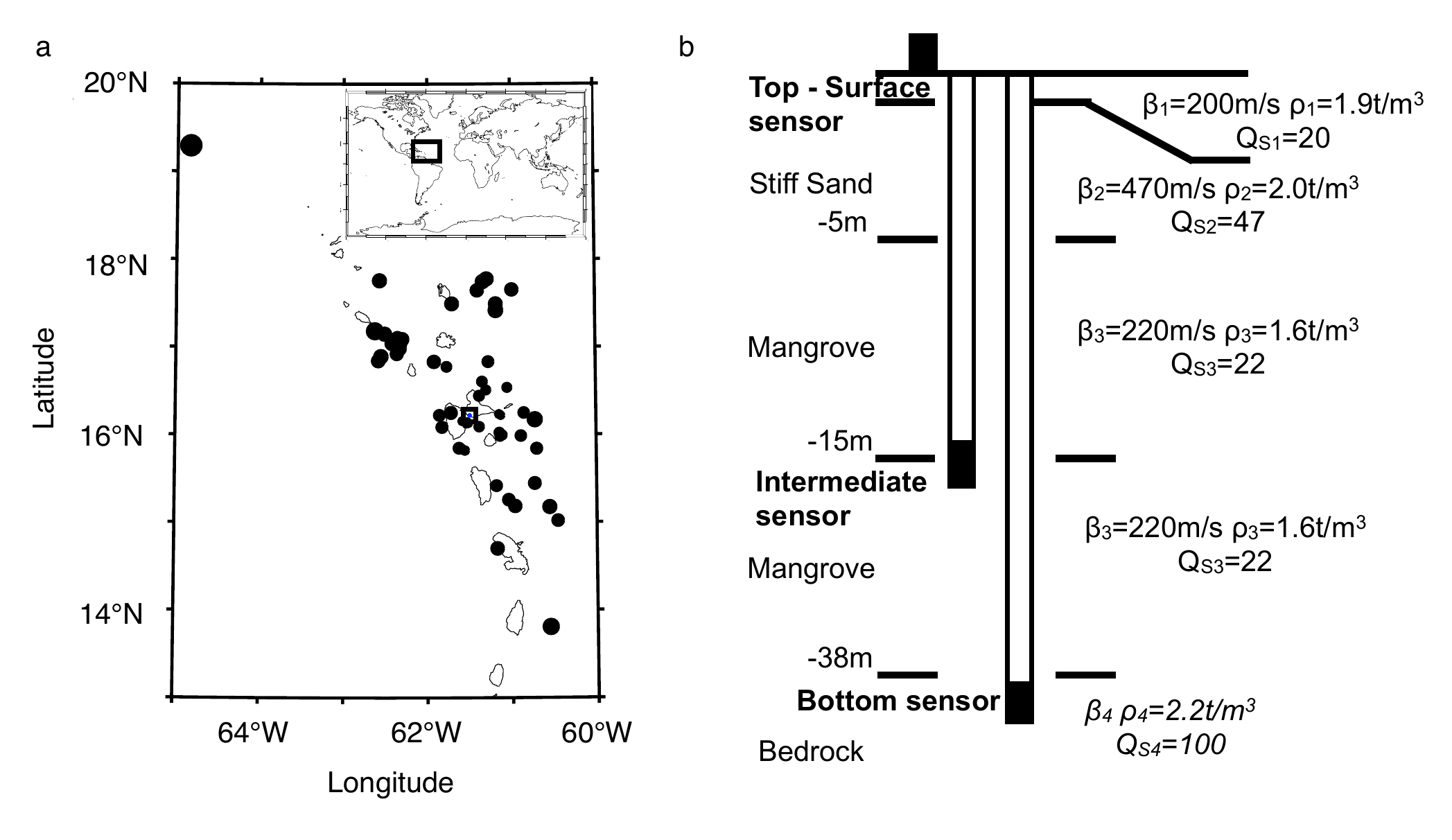}
\caption{Figure 1  }
\end{figure}

\newpage

\begin{figure}
\noindent\includegraphics[width=30pc]{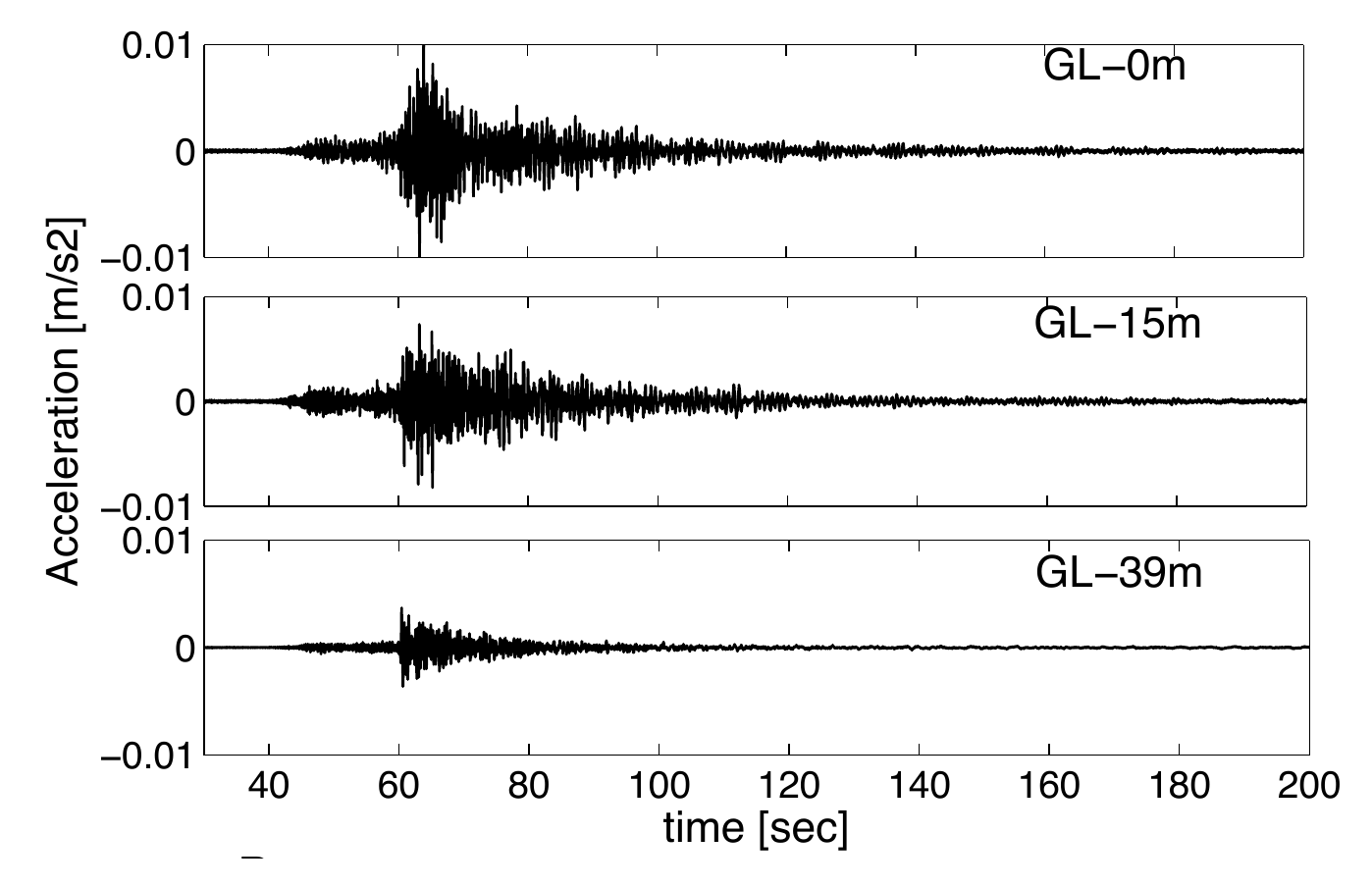}
\caption{Figure 2 }
\end{figure}

\newpage

\begin{figure}
\noindent\includegraphics[width=40pc]{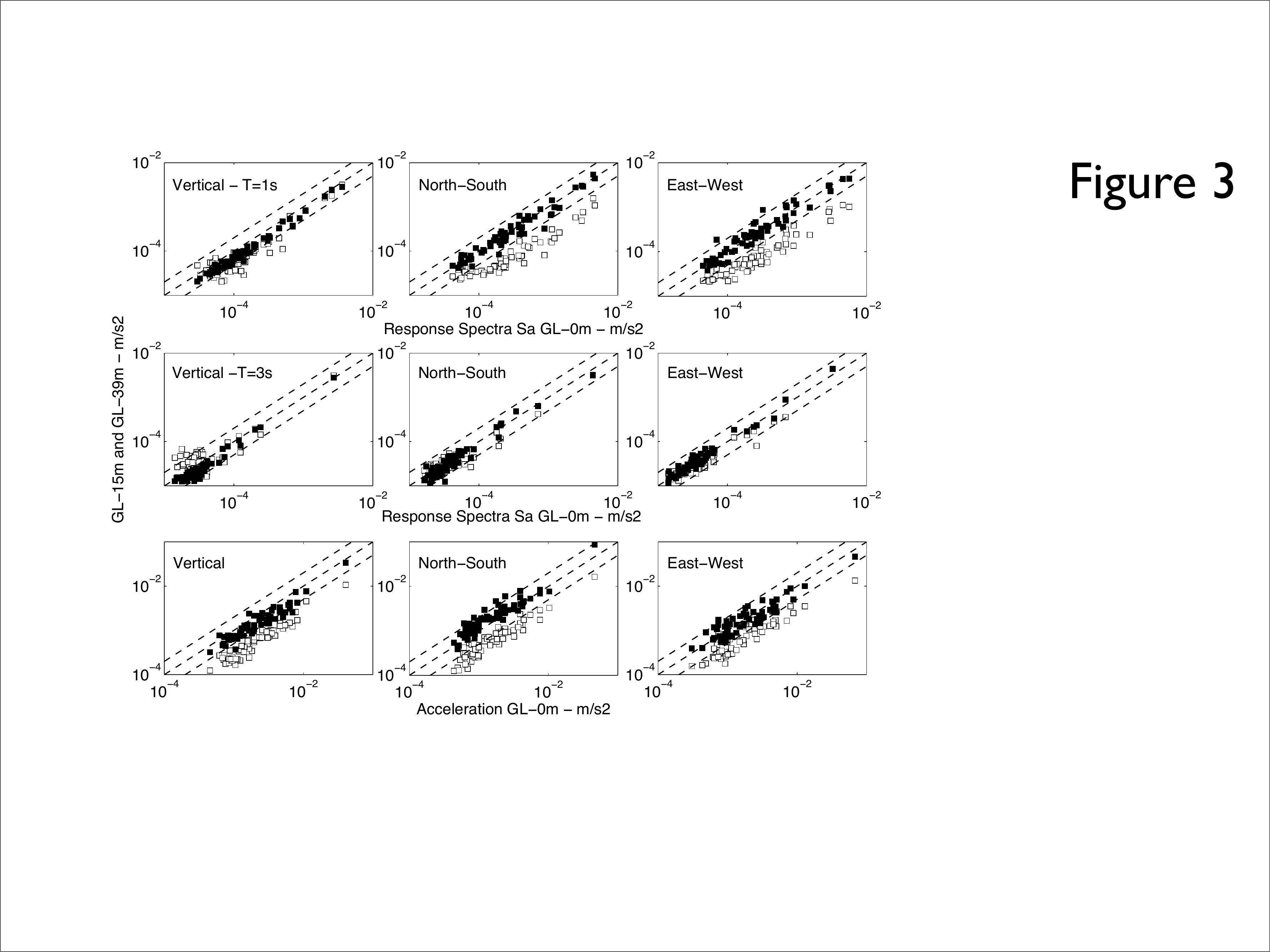}
\caption{Figure 3}
\end{figure}

\newpage

\begin{figure}
\noindent\includegraphics[width=40pc]{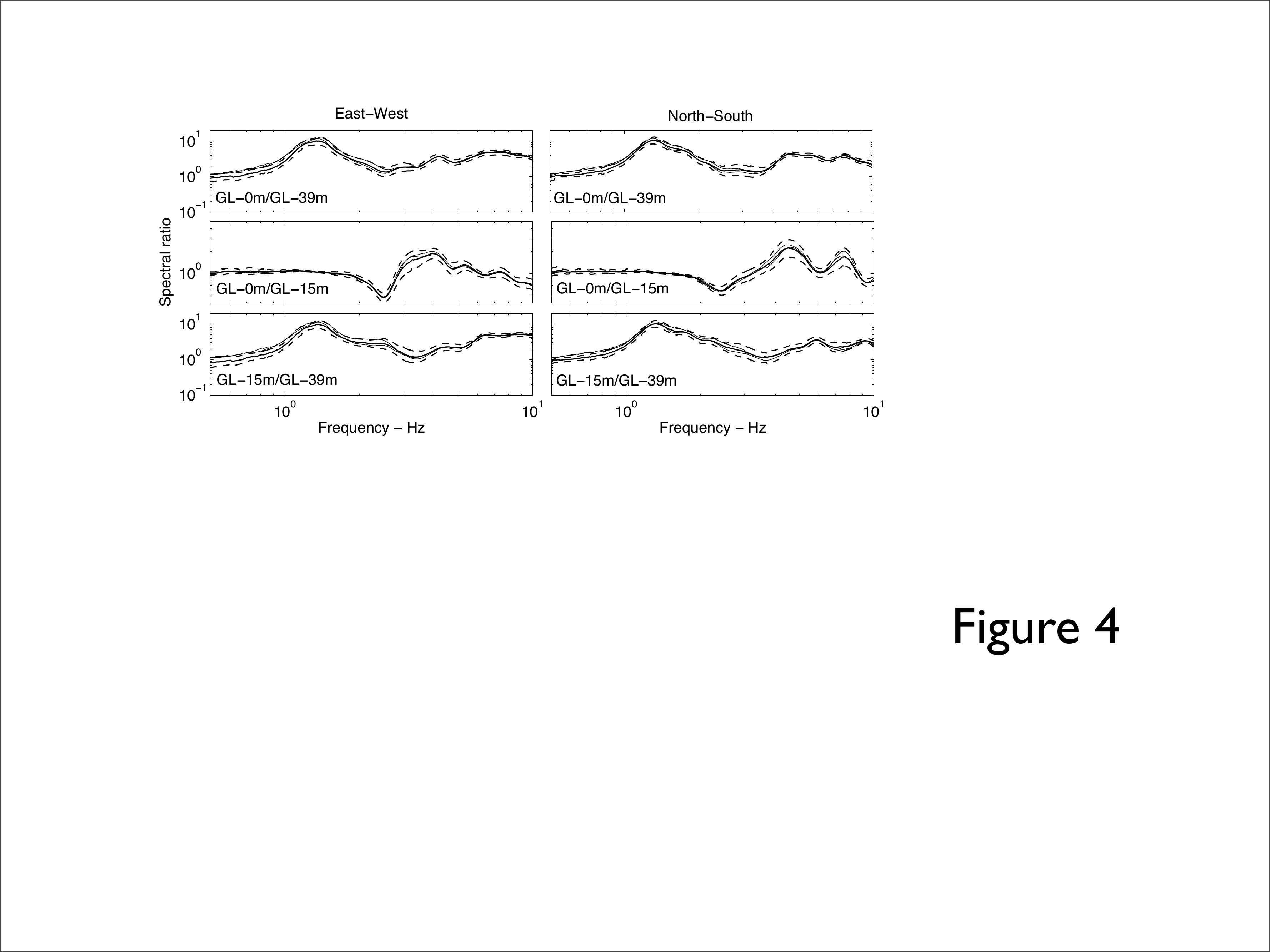}
\caption{Figure 4}
\end{figure}

\newpage

\begin{figure}
\noindent\includegraphics[width=30pc]{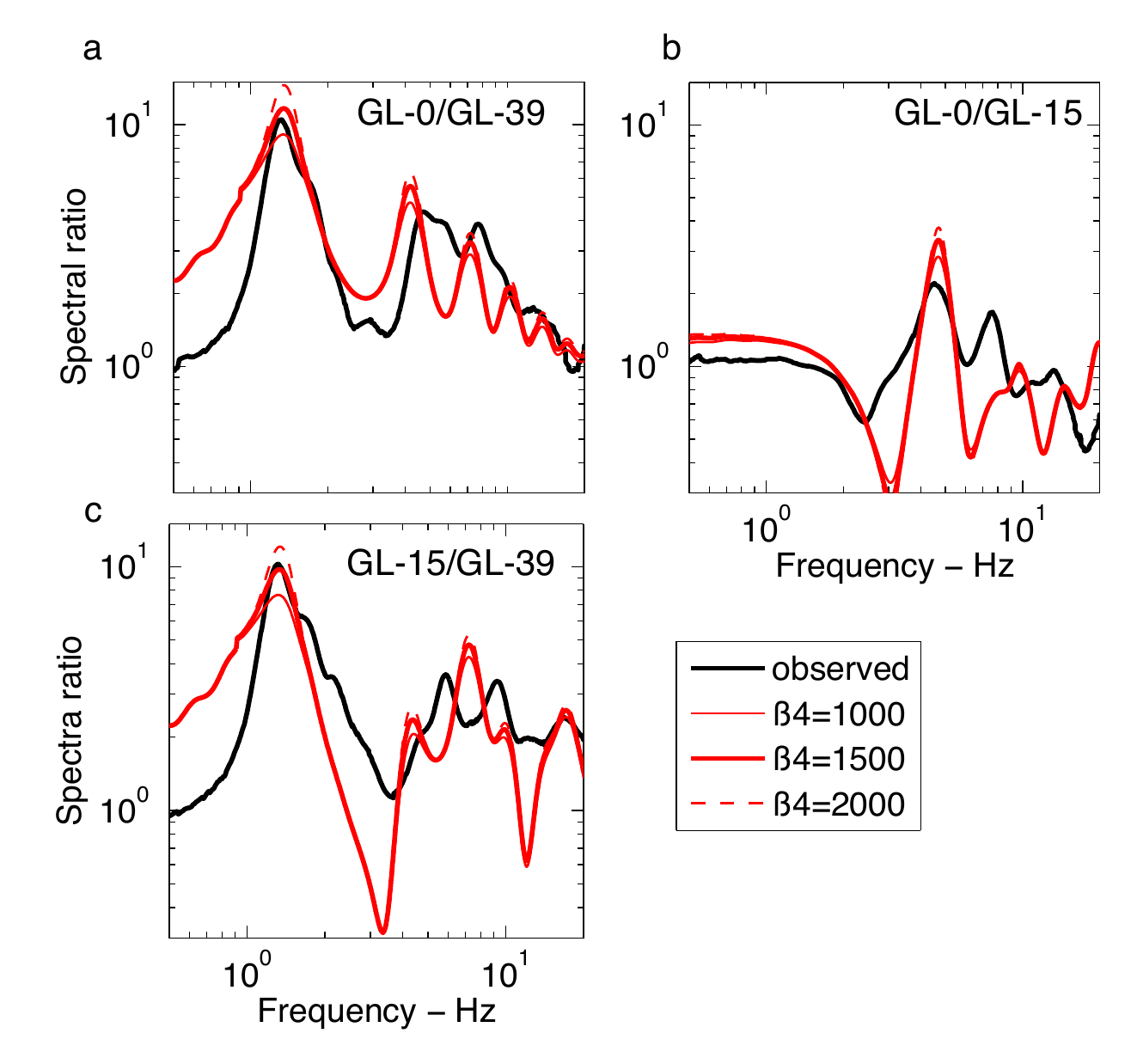}
\caption{Figure 5}
\end{figure}

 \begin{figure}
\noindent\includegraphics[width=20pc]{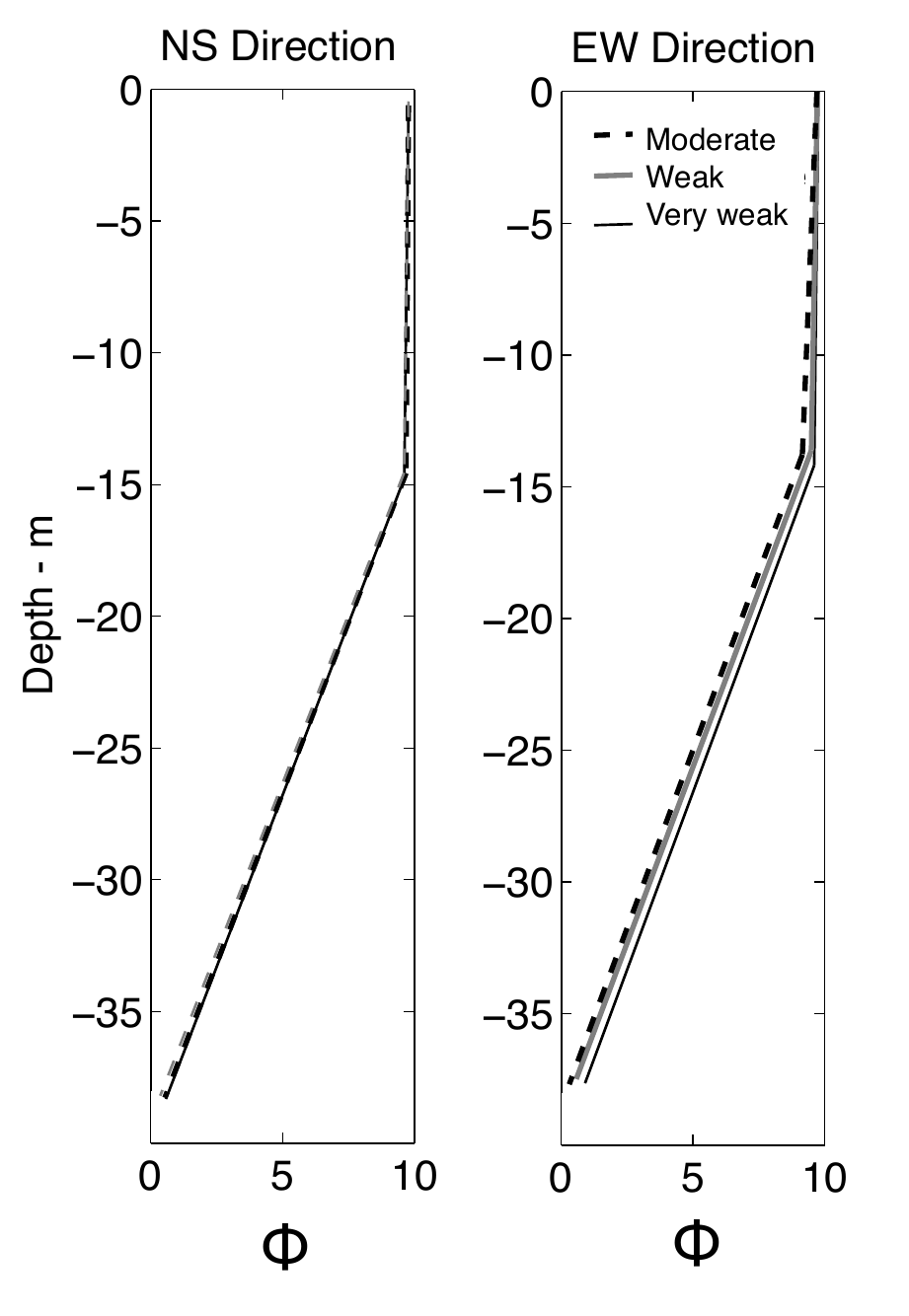}
\caption{Figure 6}
\end{figure}

 \begin{figure}
\noindent\includegraphics[width=40pc]{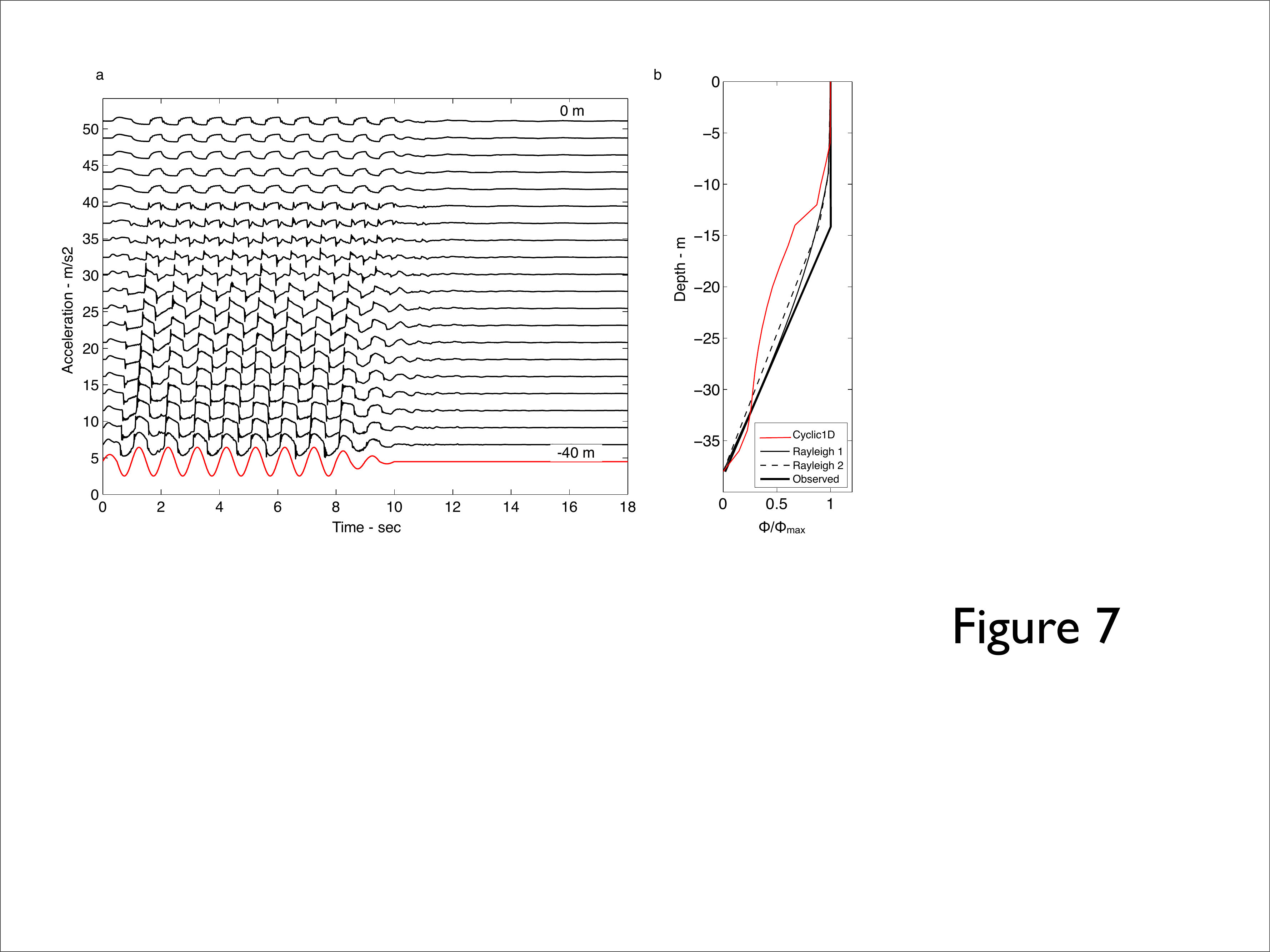}
\caption{Figure 7}
\end{figure}

\end{document}